\documentstyle[preprint,aps,epsf]{revtex}
\begin{document}
\draft

\title{\bf $J/\Psi$ suppression at SPS, RHIC and LHC energies and the
percolation of strings}

\author{E. G. Ferreiro, F. del Moral and C. Pajares}
\address{Departamento de F\'{\i}sica de Part\'{\i}culas\\
Universidade de Santiago de Compostela\\
15706 Santiago de Compostela, Spain\\}
\vskip 0.5cm
\author{N. Armesto}
\address{Departamento de F\'{\i}sica, M\'odulo C2,
Pl. Baja\\
Campus de Rabanales, Universidad de C\'ordoba\\
14071 C\'ordoba, Spain
}


\maketitle

\begin{abstract}
We study the enhancement of $c \bar c$ pair production that takes place in
central heavy ion collisions due to the formation of clusters of strings.
These clusters produce heavy flavors more efficiently due to their higher
color.
We discuss the competition between this mechanism and 
the well-known screening of color charge, which, above a critical 
string density, reduces strongly the probability of binding
these $c \bar c$ pairs to form $J/\Psi$ particles. The dependence
of $J/\Psi$ suppression on the centrality shows a peak at
both RHIC and LHC energies corresponding
to the percolation critical density.
\end{abstract}


In most of the hadronic models of multiparticle production, color strings
are exchanged between projectile and target. These strings decay afterwards
into the observed
hadrons. Color strings may be viewed as small areas $\pi r_0^2$, $r_0 \simeq$
0.2 fm,
in the transverse
space, filled with the color field created by
the colliding partons. Particles are produced via emission of $q \bar q$
pairs
in this field. 
The number of exchanged strings grows with the energy, the centrality and 
the atomic number
of the colliding
particles; thus it is natural to consider that
they start to overlap, 
forming clusters, very much like disks in continuum two-dimensional percolation
theory. At a certain critical density a large cluster appears,
which signs the percolation phase transition \cite{ref1,ref2,ref3,ref4}.

A cluster of $n$ strings 
behaves as a single string with a higher color field $\vec Q_n$,
corresponding to the vectorial sum of the color charge of each individual
$\vec Q_1$ string. The resulting color field covers the area $S_n$ of 
the cluster. As $\vec Q_n^2=( \sum_1^n \vec Q_1)^2$ \cite{ref5}, 
and the individual
string colors may be oriented in an arbitrary manner respective to one
another, the average $<\vec Q_{1i}\vec Q_{1j}>$ is zero, so
$<\vec Q_n^2>=<n \vec Q_1^2>$.

The production of $c \bar c$ pairs via tunneling effect in a 
single string is 
usually assumed to be
quite negligible compared with the production
in hard collisions,
and also negligible compared with the light flavor pair
production. However, a cluster formed by many strings has a very high
color and therefore a very large string tension which can enhance the
$c \bar c$ pair production several orders of magnitude \cite{ref5,ref6}.
This effect works in the opposite direction to the Debye screening \cite{ref7}
which
makes that above the percolation threshold the probability of binding
the $c \bar c$ pair to form a $J/\Psi$ is strongly reduced.

In this paper, we compute in a single and direct way the two effects at 
SPS, RHIC and LHC energies. 
The $J/\Psi$ suppression as an evidence \cite{ref7b} of the obtention of the
Quark Gluon Plasma has been extensively discussed in the last years 
\cite{ref1,ref2,ref7c}. 
We would like to answer whether the abnormal
$J/\Psi$ suppression configuration pattern for central Pb-Pb collisions
at SPS energies is going to be modified at RHIC and/or LHC energies.
Recently, a picture of the $J/\Psi$ creation via $c \bar c$
coalescence (recombination) has been subsequently developed within different
model formulations \cite{ref8,ref9,ref10,ref11,ref12}. Charmonium
states are assumed to be created at the hadronization stage of the reaction,
being formed due to the coalescence of $c$ and $\bar c$,
which were produced by some primary mechanism
(hard parton collisions or statistical decay of a fireball)
at the initial
stage. In these approaches, it is found, instead of suppression, an
enhancement of the $J/\Psi$ production at RHIC and LHC energies. 

On the contrary, in our approach we find that the $J/\Psi$ suppression 
survives at RHIC and LHC energies and only a relative enhancement can
be seen close to the percolation critical density.
We are aware of the existence of other effects like shadowing and the
existence of coherence length \cite{ref13,ref14}, which we do not take
into account. These effects at high energies are very important for
open charm production. However, in the central rapidity region for the case
of $J/\Psi$ production, they are expected to be of little importance
(less than 5-10 \%). Also, we
do not pay attention to whether there is one or more steps in the
suppression pattern corresponding to the $J/\Psi$ directly
produced in the collisions or considering also the suppression of other
resonances which decay into $J/\Psi$ \cite{ref15}. The inclusion of
this possibility could be done easily but this 
is not the main goal of our study: our computation 
is focused on the
               qualitative behavior produced by the
               competition of suppression and enhancement coming from
               the same origin, the high color fields created by the
               overlapping strings exchanged in the collision.

Let us start by considering the extension of the Schwinger formula
\cite{ref5,ref6} for the production of $q- \bar q$ pairs of  mass
$m_j$ in a uniform color field with charge $g_j$, per unit space-time
volume

\begin{equation}{dN_{q- \bar q} \over dy} =  {1 \over 8 \pi^3}
\int_0^\infty d \tau \tau\int d^2x_T
|g_j E|^2\sum_{n=1}^\infty 
{1 \over n^2} 
\exp \Big (  -{\pi n m_j^2\over |g_j E|}\Big )\ \ .
\label{ec1}\end{equation}

The strings form clusters, 
each of them with a constant color field $E_i=Q_i/S_i$,
where $Q_i$ and $S_i$ correspond to the cluster color charge and the cluster area,
respectively. Hence, the integral in transverse space in the above equation
transforms into a sum of areas $S_i$ of the clusters.

On the other hand, we take into account the evolution of the field and the charge
with the decay of the cluster,

\begin{equation} E_i=E_{i0}{ 1 \over (1+ {\tau \over \tau_{i0}})^2} \qquad , \qquad 
Q_i= Q_{i0} { 1 \over (1+ {\tau \over \tau_{i0}})^2}\ \ ,
\end{equation}
where $\tau \sim 1/ \sqrt{E_{i0}}$.

In this way, equation (\ref{ec1}) transforms into the equation
\begin{equation}{dN_{q- \bar q} \over dy} \propto  
\sum_{i=1}^M Q_{i0}\int_0^\infty dx x 
{ 1 \over (1+ x)^4}\sum_{n=1}^\infty 
{1 \over n^2} \exp \Big [  -{\pi n m_j^2\over |g_j E_{i0}|} 
(1+ x)^2\Big ]\ \ ,
\label{ec3}\end{equation}
where $M$ is the total number of clusters.

The charge and the field of each cluster before the decay, $Q_{i0}$ and 
$E_{i0}$,
depend 
on the number $n_i$ of strings and the area $S_1$ of each individual string
that comes into the cluster, as well as on 
the total area of the cluster $S_i$ \cite{ref4},
\begin{equation}Q_{i0}= \sqrt{ n_i S_i \over S_1} Q_{10}\qquad , \qquad E_{i0}=
{Q_{i0} \over S_i}=\sqrt{ n_i S_1 \over S_i}E_{10}\ \ .
\label{ec4}\end{equation}

Notice that if the strings are just touching each other, $S_i=n_i S_1$ 
and $Q_{i0}=nQ_{10}$,
$E_{i0}=E_{10}=0.2$ GeV$^2$, so 
the strings are independent each other. On the contrary, if they
fully overlap, $S_i=S_1$ and $Q_{i0}=\sqrt {n_i} Q_{10}$ and $E_{i0}=\sqrt {n_i}E_{10}$,
so the field reaches its maximum strength.

In order to compute 
$J/\Psi$ production, we proceed as reference \cite{ref16}. We shall
consider that when $\nu$ collisions occur they occupy an interaction volume $V(\nu )$
characterized by a transverse area $A(\nu )$ and a mean longitudinal length $L(\nu )$,
\begin{equation}V(\nu ) \equiv A(\nu )L(\nu )\ \ .
\end{equation}

The density of strings is
\begin{equation} \rho_s= {N_s \over A(\nu )L(\nu )}\ \ ,
\end{equation}
where $N_s$ is the number of strings. We introduce the survival probability $P_s(\nu )$,
to take care of the $J/\Psi$ absorption \cite{ref17} in the usual form
\begin{equation}P_s(\nu )=\exp (-L(\nu ) \rho_s \sigma)\ \ ,
\end{equation}
$\sigma$ being the pre resonant $c\bar c$-nucleon absorption cross section.
This corresponds to the usual picture of the conventional mechanism
proportional to $A$ plus absorption by nuclear matter.

Next, 
we include the probability of quark-gluon plasma formation in a very simple
continuum two-dimensional percolation model. In an event with $\nu$ elementary 
collisions,
there is a probability $P_{np}(\nu )$ of percolation not to occur, so a 
probability $1-P_{np}(\nu )$ of percolation to occur. It is assumed that the 
$J/\Psi$ is formed only in events in which there is not percolation.

The probability of having an 
infinite cluster in terms of $\eta = \pi r_0^2 N_s /A(\nu )$,
$N_s=2\nu$, is 
\begin{equation}P_{perc}= \theta( \eta -\eta_c)\ \ , \end{equation}
where 
the percolation threshold has been computed by several authors to be in the 
range 1.12-1.17. Because of the finite size of the colliding system, the theta
function will be approximate by a smoother function. Indeed \cite{ref16}
\begin{equation}P_{perc}= 1/(1+\exp (-(\eta -\eta_c)/a)) \end{equation}
approximates 
very well the percolation behavior with $\eta_c=1.15$ and $a=0.04$.

Let $N(\nu )$ be the total number of events with $\nu$ elementary
collisions, and $N_c(\nu )$ the number of events when a rare event C
(an event is called rare
if, in good approximation, it only occurs once in each collision) 
occurring \cite{ref18},
\begin{equation}N_C(\nu )= \alpha_C \nu N(\nu )\ \ , \end{equation}
where $\alpha_C$ is the probability of event $C$ occurring in an elementary
collision. 
Corrections to this formula due to several $c \bar c$ pairs in a collision 
are proved to be small \cite{ref19}.
Now we can apply the formulae developed before to $J/\Psi$ and Drell-Yan
production.

The probability to find a Drell-Yan event when there are $\nu$ elementary
collisions is
\begin{equation} P(DY|\nu )= \alpha_{DY} \;\nu\ \ , \end{equation}
while for $J/\Psi$ production one has
\begin{equation} P(J/\Psi |\nu )= \alpha_{J/\Psi}\;\nu\; e^{-\sigma \eta /  \pi r_0^2}\;
{1 \over \exp ({\eta-\eta_c \over a})+1}\ \ , \end{equation}
and the ratio $R$ of $J/\Psi$ to D-Y events is
\begin{equation}R=k \exp (-\sigma \eta /  
\pi r_0^2)/[\exp ({(\eta-\eta_c) /a})+1]\ \ . 
\label{ec10}\end{equation}

However, as we said before, the probability of $c \bar c$ production increases with
$\eta$. In order to include this fact, we multiply the expression (\ref{ec10}) by
\begin{equation}P_{c\bar c}(\eta )/P_{c\bar c}(\eta_{SPS} )\label{ec11}\end{equation}
assuming that the leading $c \bar c$ pair production for central Pb-Pb collisions at SPS
energies is given by the 
Schwinger mechanism, expression (\ref{ec3}) and (\ref{ec4}).
$P_{c\bar c}(\eta )$ is the probability of $c\bar c$ production at a given $\eta$
value and is normalized to the corresponding probability for central Pb-Pb collisions
at SPS energies in such a way that the correction factor is just one at SPS energies.
Similar evaluations have been
extensively used in strangeness production giving reasonable
results \cite{ref20,ref21,ref22,ref23}.

We use formulae
(\ref{ec10}) and (\ref{ec11}) to compute the ratio $J/\Psi/DY$ at RHIC and LHC
as a function of the transverse energy $E_T$ and the average interaction distance $L$, which
are two usual variables to measure the degree of centrality of the collisions. At SPS we used
the results of reference \cite{ref16} directly obtained from formula (\ref{ec10}).
To compute the ratio (\ref{ec11}) we use formula (\ref{ec3}) and (\ref{ec4}) where the clusters
of strings and their areas are evaluated generating localized strings is impact parameter plane
by means of a Monte-Carlo code based on the Quark Gluon String Model.

\begin{figure}
  \centering\leavevmode
  \epsfxsize=5in\epsfysize=5in\epsffile{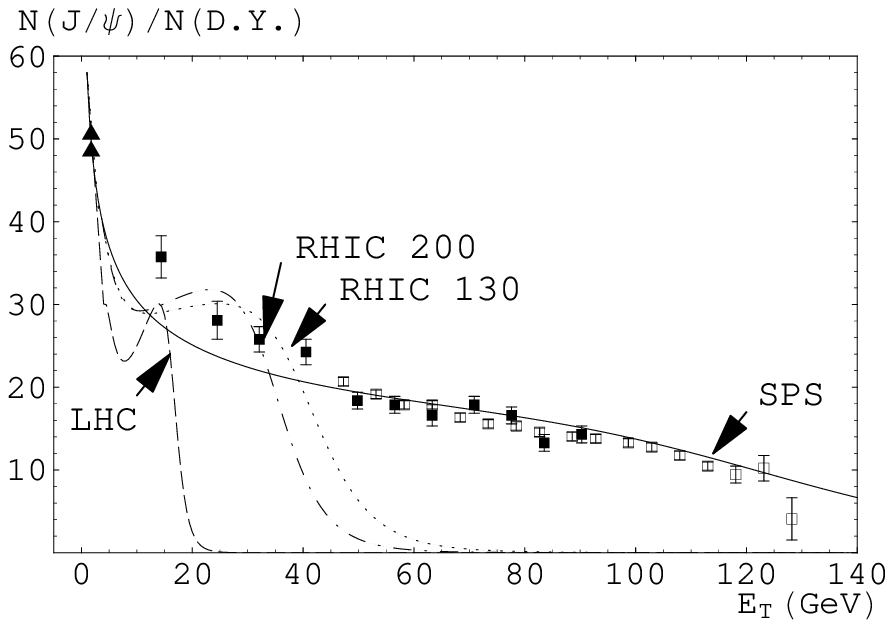}
\caption{Ratio of $J/\Psi$ to Drell-Yan events as predicted by (\ref{ec10}) and
(\ref{ec11}) 
for Pb-Pb collisions at SPS energies (solid line),
Au-Au collisions at RHIC energies (130 GeV/n) (dotted line),
Au-Au collisions at RHIC energies (200 GeV/n) (dashed-dotted line)
and Pb-Pb collisions at LHC energies (dashed line)
as a function of $E_T$. Arrows mark the percolation critical points and
experimental data for SPS are from NA50 Collaboration [8,26].}
\label{figure1}
\end{figure}

\begin{figure}
  \centering\leavevmode
  \epsfxsize=5in\epsfysize=5in\epsffile{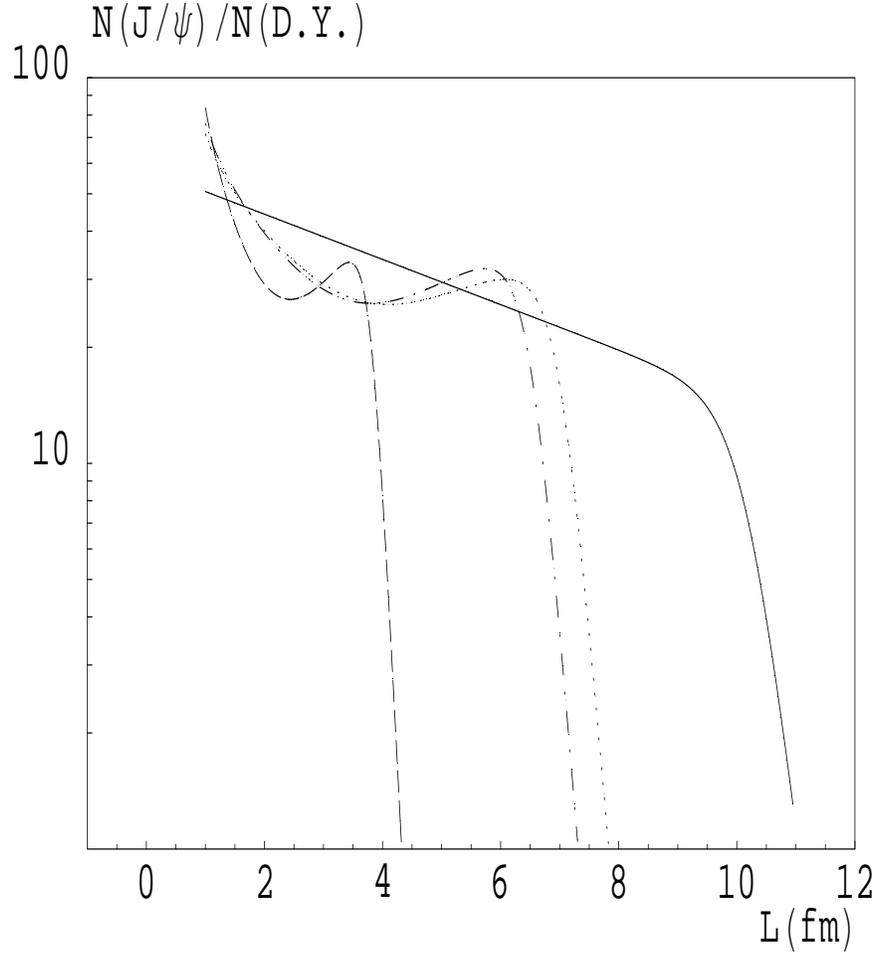}
\caption{Ratio of $J/\Psi$ to Drell-Yan events as predicted by (\ref{ec10}) and
(\ref{ec11}) 
for Pb-Pb collisions at SPS energies (solid line),
Au-Au collisions at RHIC energies (130 GeV/n) (dotted line),
Au-Au collisions at RHIC energies (200 GeV/n) (dashed-dotted line)
and Pb-Pb collisions at LHC energies (dashed line)
as a function of $L$.}
\label{figure2}
\end{figure}

In figures \ref{figure1} and \ref{figure2} our results for the ratio $J/\Psi/DY$
as a function of $E_T$ and $L$ are plotted. 
It is seen that at RHIC and LHC energies an increase appears just
before the percolation critical point which is marked by arrows in fig. \ref{figure1}.
At higher centralities the ratio $J/\Psi/DY$ drops sharply. The same pattern is seen in 
figure \ref{figure2} where it is used $L$ instead of $E_T$.

\begin{figure}
  \centering\leavevmode
  \epsfxsize=5in\epsfysize=5in\epsffile{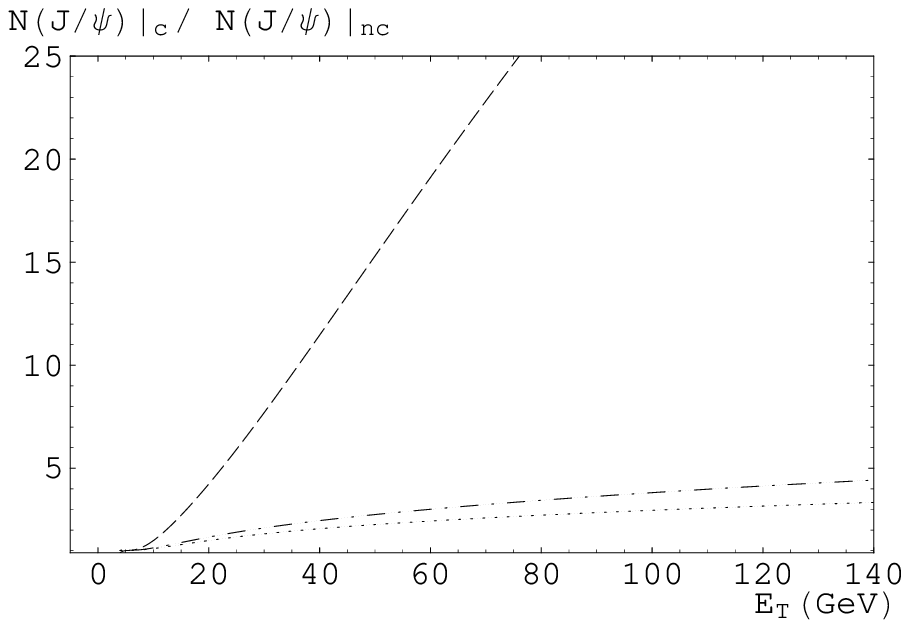}
\caption{Ratio of predictions for $J/\Psi$ suppression with and without clustering of
strings, equation (\ref{ec11}), 
for Au-Au collisions at RHIC energies (130 GeV/n) (dotted line),
Au-Au collisions at RHIC energies (200 GeV/n) (dashed-dotted line)
and Pb-Pb collisions at LHC energies (dashed line)
as a function of $E_T$.}
\label{figure3}
\end{figure}

In fig. \ref{figure3} we plot the correction ratio (\ref{ec11}) as a function of $E_T$ for
RHIC and LHC energies. Notice that in spite of becoming very large for $E_T > 40$ GeV at LHC
its effect is quite negligible in the production of $J/\Psi$ in that range. Only, just before
the percolation critical points, the large $c\bar c$ production gives rise to noticeable 
effects. We are aware of some uncertainties of our evaluations, related to the 
normalization used in (\ref{ec11}). However we have checked that a
change in  (\ref{ec11}) by a factor 2 would
not kill the reported effect.

Our results 
are different from other models \cite{ref8,ref9,ref10,ref11,ref12} based
on statistical 
hadronization of charm quarks without color screening scenario, which 
predict 
$J/\Psi$ enhancements at RHIC and LHC energies. 
In those models, the $J/\Psi$ is produced at the end of the process, due to
hadronization of the quarks, and this is the main reason of the enhancement.
In our approach, we cannot exclude the formation of some $J/\Psi$ particles
after the hadronization of the plasma, but this production will be 
proportional to the number of $c\bar c$ pairs computed by means of the ratio 
(14) plotted in fig. \ref{figure3}. At RHIC energies this ratio is very small
compared to the suppression. The same is true at LHC energies in the range
$20 \le E_T \le 100$ GeV. Therefore, we expect that our general pattern of
$J/\Psi$ suppression will not be modified by the hadronization of the plasma.
This effect would be important only at LHC energies for very high $E_T$.
Notice that the enhancement of $c$ quarks is due in our case to the higher
tension of the string clusters, while in those models is due to statistical
exponentials. 
These clear differences between different predictions
               make the verification more relevant.

Finally, let us summarize our main results. We have computed the enhancement of
$c\bar c$ pair production in heavy ion collisions due to the formation of
clusters of string with high color. This enhancement is not enough to prevent
the suppression of the $J/\Psi$ once the percolation critical density is
reached and the Quark Gluon Plasma is formed. The dependence of the $J/\Psi$ 
suppression with the centrality shows a peak located just before the
percolation critical density. 

\acknowledgments

This work has been done under contracts AEN99-0589-C02 of CICYT
of SPAIN, PGID-TOOPXI20613PN of
Xunta de Galicia. F. del M. thanks Xunta de Galicia for a fellowship. 
N. Armesto thanks Univ. de C\'ordoba for financial support. We thank
M. A. Braun for useful discussions.

\begin{references}

\bibitem{ref1} N.~Armesto, M.~A.~Braun, E.~G.~Ferreiro and C.~Pajares, Phys.\ Rev.\ Lett.\ {\bf 77},
3736 (1996).

\bibitem{ref2} M.~Nardi and H.~Satz, Phys.\ Lett.\ {\bf B442}, 14 (1998); H.~Satz, Nucl.\ Phys.\ 
{\bf A642}, 130c (1998).

\bibitem{ref3} M.~A.~Braun and C.~Pajares, Eur.\ Phys.\ J.\ {\bf C16}, 349 (2000);
M.~Braun, C.~Pajares and J.~Ranft, Int. J. of Mod. Phys. {\bf A14}, 2689 (1999).

\bibitem{ref4} M.~A.~Braun and C.~Pajares, Phys.\ Rev.\ Lett.\ {\bf 85},
4864 (2000); M.~A.~Braun, F.~del~Moral and C.~Pajares, hep-ph/0105263.

\bibitem{ref5} T.~Matsui, MIT preprint MIT-CTP1510 (1987).

\bibitem{ref6} J.~Schwinger, Phys.\ Rev.\ {\bf 82}, 664 (1951); T.~S.~Biro, H.~B.~Nielsen
and J.~Knoll, Nucl.\ Phys.\ {\bf B245}, 449 (1984); A.~Bialas and W.~Czyz, 
Nucl.\ Phys.\ {\bf B267}, 242 (1986).

\bibitem{ref7} T.~Matsui and H.~Satz, Phys.\ Lett.\ {\bf B178}, 416 (1986).

\bibitem{ref7b} M. C. Abreu {\it et al.}, NA50 Collaboration, Phys. Lett. {\bf
B477}, 28 (2000).

\bibitem{ref7c} J. Blaizot and J.-Y. Ollitrault, Phys. Rev. Lett. {\bf 77},
1703 (1996); 
J. Blaizot, M. Dinh and J.-Y. Ollitrault, Phys. Rev. Lett. {\bf 85}, 4012
(2000);
C. Y. Wong, Phys. Rev. Lett. {\bf 76}, 196 (1996); Phys. Rev. {\bf C55}, 2621
(1997);
N. Armesto, A. Capella, E. G. Ferreiro, Phys. Rev. {\bf C59}, 395 (1999);
A. Capella, E. G. Ferreiro and A. B. Kaidalov, Phys. Rev. Lett. {\bf 85},
2080 (2000);
A. Capella, A. B. Kaidalov and D. Sousa, nucl-th/0105021 (2001);
J. Geis, C. Greiner, E. L. Bratkovskaya, W. Cassing and U. Mosel, Phys.\
Lett.\ {\bf B447}, 31 (1999);
J. Qin, J. P. Vary and X. Zhang, nucl-th/0106040 (2001).


\bibitem{ref8} R.~L.~Thews, M.~Schroedter and J.~Rafelski, Phys.\ Rev.\ {\bf C63},
054905 (2001).

\bibitem{ref9} M.~I.~Gorenstein, A.~P.~Kostyuk, L.~McLerran, H.~St\" ocker and
W.~Greiner, Brookhaven preprint BNL-NT-00/27 (2000).

\bibitem{ref10} 
P.~Braun-Munzinger and J.~Stachel, Phys.\ Lett.\ {\bf B490}, 196 (2000).

\bibitem{ref11} P.~Levai, 
T.~S.~Biro, P.~Csizmadia, T. Cs\"org\"o and J.~Zimanyi, J. Phys {\bf 27},
703 (2001).

\bibitem{ref12} S.~Kabana, hep-ph/0004138.

\bibitem{ref13} N.~Armesto, M.~A.~Braun, A.~Capella, C.~Pajares and C.~A.~Salgado,
Nucl.\ Phys.\ {\bf B509}, 357 (1998); C.~A.~Salgado, hep-ph/0105231.

\bibitem{ref14} B.~Z.~Kopeliovich, A.~Tarasov and J.~H\" ufner, hep-ph/0104256;
Y.~B.~He, J.~H\"ufner and B.~Z.~Kopeliovich, Phys.\ Lett.\ {\bf B477}, 93 (2000).

\bibitem{ref15} H.~Satz, Nucl.\ Phys.\ {\bf A661}, 104c (1999).

\bibitem{ref16} J.~Dias de Deus, R.~Ugoccioni and A.~Rodrigues, Eur.\ Phys.\ J.\ {\bf C16}, 537 (2000).

\bibitem{ref17} A.~Capella, J.~A.~Casado, C.~Pajares, A.~V.~Ramallo and J.~Tran Thanh Van,
Phys.\ Lett.\ {\bf B206}, 354 (1988); C.~Gerschel and J.~H\" ufner, Phys.\ Lett.\ {\bf B207}, 253
(1988).

\bibitem{ref18} J.~Dias de Deus, C.~Pajares and C.~A.~Salgado, Phys.\ Lett.\ {\bf B407}, 335 (1997);
{\bf B409}, 474 (1997).

\bibitem{ref19} J.~Dias de Deus and C.~Pajares, Phys.\ Lett.\ {\bf B442}, 395 (1998).

\bibitem{ref20} N.~S.~Amelin, M.~A.~Braun and C.~Pajares, Phys.\ Lett.\ {\bf B306}, 312 (1993);
Z.\ Phys.\ {\bf C63}, 507 (1994).

\bibitem{ref21} N.~S.~Amelin, N.~Armesto, C.~Pajares and D.~Sousa, hep-ph/0103060.

\bibitem{ref22} Y.~M.~Shabelski, Surveys in High Energy Physics {\bf 9}, 1 (1995);
P.~Koch and U.~Heinz 
in ``Quark Gluon Plasma Signatures'', Editions Frontieres (1991).

\bibitem{ref23} M.~Bleicher, M.~Belkacem, S.~A.~Bass, S.~Soff and H.~St\" ocker, 
Phys.\ Lett.\ {\bf B485}, 133 (2000).

\bibitem{ref24} M.~C.~Abreu et al. (NA50 Collab.), Phys.\ Lett.\ {\bf B409}, 474 (1997);
C.~Cicalo (NA50 Collab.), Nucl.
                  Phys. {\bf A661}, 93c (1999).
 

\end{references}
\end{document}